\documentclass{osa-article}

\journal{osac}

\articletype{Research Article}

\begin{document}

\title{Polarization-resolved broadband time-resolved optical spectroscopy for complex materials: application to the case of MoTe$_2$ polytypes}

\author{Michele Perlangeli,\authormark{1} Simone Peli,\authormark{2} Davide Soranzio,\authormark{1} Denny Puntel,\authormark{1} Fulvio Parmigiani,\authormark{1,2,3} and Federico Cilento\authormark{2,*}}

\address{\authormark{1}Dipartimento di Fisica, Università degli Studi di Trieste, 34127 Trieste, Italy\\
\authormark{2}Elettra-Sincrotrone Trieste S.C.p.A., 34149 Basovizza, Italy\\
\authormark{3}International Faculty, University of Cologne, Albertus-Magnus-Platz, 50923 Cologne, Germany}

\email{\authormark{*}federico.cilento@elettra.eu}


\begin{abstract}
Time-resolved optical spectroscopy (TR-OS) has emerged as a fundamental spectroscopic tool for probing complex materials, to both investigate ground-state-related properties and trigger phase transitions among different states with peculiar electronic and lattice structures. We describe a versatile approach to perform polarization-resolved TR-OS measurements, by combining broadband detection with the capability to simultaneously probe two orthogonal polarization states. This method allows us to probe, with femtoseconds resolution, the frequency-resolved reflectivity or transmittivity variations along two mutually orthogonal directions, matching the principal axis of the crystal structure of the material under scrutiny. We report on the results obtained by acquiring the polarization-dependent transient reflectivity of two polytypes of the MoTe$_2$ compound, with 2H and 1T' crystal structures. We reveal marked anisotropies in the time-resolved reflectivity signal of 1T'-MoTe$_2$, which are connected to the crystal structure of the compound. Polarization- and time- resolved spectroscopic measurements can thus provide information about the nature and dynamics of both the electronic and crystal lattice subsystems, advancing the comprehension of their inter-dependence, in particular in the case of photoinduced phase transitions; in addition, they provide a broadband measurement of transient polarization rotations.
\end{abstract}

\section{Introduction}
In the last years, time-resolved optical measurements have been used to shed light on a large variety of open issues on the physics of complex and strongly-correlated materials \cite{Giannetti_Review}. Time-domain measurements have been widely used to disentangle, by their different timescales, the degrees of freedom of a system that at equilibrium are intertwined, thus advancing the comprehension of ground state properties of solid state compounds \cite{Liu2008}. Moreover, photoexcitation at high fluence ($\approx$0.2-2 mJ/cm$^2$) makes it possible to trigger phase transitions among different states of a sample, often not attainable under equilibrium conditions, obtaining a true ultrafast control over the functional properties of materials \cite{Tomeljak2009}. The advent of spectroscopic measurements, making use of broadband, supercontinuum pulses \cite{Polli2007,Leonard2007,Cilento2010,Chergui2012a}, helped to put on a more solid base the interpretation of the reflectivity or transmittivity transients \cite{Megerle2009}, by connecting the measured quantities to specific modifications of the dielectric function and the underlying band structure \cite{Giannetti2011}. The progress of laser and instrumentation technology improved this approach, so that weak transient signals as originated by a gentle excitation of samples can nowadays be revealed \cite{Preda2016,Chergui2012b}. Here, we describe a novel approach combining broadband detection with the ability to probe simultaneously two polarization components of the probe pulse \cite{Hiramatsu2015}. When the crystal axes of the compound under scrutiny are oriented along these components, it is possible to probe the lattice / structural degree of freedom. As we will show in the following, this fact is important for materials possessing anisotropic optical properties. Under equilibrium conditions, the anisotropy can be ascribed either to the symmetry of the crystal lattice, or to a specific electronic ordering (among which nematicity is an example). Anisotropies can also emerge as a result of photo-excitation. In this situation, the possibility to probe simultaneously the optical properties along two crystal axes can be beneficial for the comprehension of the symmetry-breaking photoinduced phase transitions. Simultaneous detection is indeed fundamental when exploring the evolution of the time-resolved optical properties as a function of external parameters like temperature, fluence or applied pressure. It ensures that the same experimental conditions, such as pump-probe overlap, sample aging, and intensity fluctuations, are shared, allowing one to reveal even the weakest polarization anisotropies. Hence, our approach opens the possibility to track in the time-domain the evolution of the electronic and lattice properties, along with their symmetry. As an example, we present the results achieved by investigating the out-of-equilibrium reflectivity variation of two MoTe$_2$ polymorphs, with 2H and 1T' structure. They differ in the electronic structure (semiconducting and semimetallic respectively) and in the lattice structure (hexagonal and monoclinic, respectively) \cite{Wilson1969}. As our results show, these differences affect the time-resolved reflectivity signal in the time domain, in the spectral domain, and in the symmetry of these signals. Information in such three domains eases to disclose the ground state properties of complex materials.

\section{Symmetry considerations}
The symmetry of the optical properties of a crystal is influenced by the spatial symmetry of its crystal lattice, that determines the form of the linear susceptibility tensor $\chi^{(1)}$. The seven possible crystal systems that a crystal lattice can assume determine the symmetry of the linear optical properties \cite{BOYD}. The natural reference system of choice to express $\chi^{(1)}$ is the one of the cartesian crystal axes x, y, z. In this notation, $\chi^{(1)}$ is a diagonal tensor, with equal components, for the cubic lattice system. Hence, in this situation the cartesian axes coincides with the principal dielectric axes. For tetragonal, rhombohedral and hexagonal lattice systems $\chi^{(1)}$ displays two different eigenvalues; the one describing the xy plane properties is doubly-degenerate. This defines uniaxial crystals. Finally, $\chi^{(1)}$ for triclinic, monoclinic, and orthorhombic structures displays three different eigenvalues, leading to biaxial crystals. Among these last structures, a difference is that $\chi^{(1)}$ is diagonal in the crystalline axis representation only for the orthorombic lattice. For triclinic and monoclinic lattices, diagonalization leads to new principal dielectric axes differing by a rotation with respect to the cartesian crystalline axes. Such a variety emphasizes the importance to systematically extend the measurement of the time-resolved optical properties including the symmetry information. Indeed, the vast majority of crystalline compounds are birefringent, and isotropic materials only constitute a special case.

\section{Experimental setup}
Our development aims at joining broadband photon energy detection with the possibility to measure simultaneously the optical properties along two crystalline axes. The optical setup is sketched in Fig. 1a). The driving laser system is a Ti:Sapphire regenerative amplifier (Coherent RegA), delivering 50 fs pulses at 800 nm, at a repetition rate of 250 kHz. The energy/pulse available is 6-7 $\mu$J. About 1 $\mu$J/pulse is used for generating a single-filament \cite{Bradler2009,Lu2014,Dubietis2017} stable supercontinuum beam, extending from 400 nm to the near-IR, which is used as a probe. The spectrum is shown in Fig. 1b) for the two simultaneously-probed components. The residual seed at 800 nm is suppressed by a 1$\%$ T @ 800 nm mirror (by Eksma Optics). The supercontinuum probe beam polarization is oriented by a polarizer at 45 degrees with respect to the horizontal. The orientation of samples is pre-determined by ex-situ Low Energy Electron Diffraction (LEED). By aligning the cartesian axes of samples along the horizontal and vertical directions, the horizontally- and vertically-polarized components of the supercontinuum beam will each probe a single principal axis. This situation is depicted in Fig. 1c). After the interaction with the sample, the probe beam is directed to a Wollaston polarizer (made of calcite by Thorlabs, and with total separation angle of 20 degrees). This single optical element carries out simultaneously two tasks, namely: i) splitting the main beam into its vertical and horizontal components; ii) dispersing the spectra on the detectors. In our setup, detector '1' is illuminated by the vertically-polarized component, while detector '2' probes the horizontally-polarized component. Interferential and colored filters have been used to perform the photon energy calibration: the dispersion characteristics of the Wollaston polarizer, as provided by the manufacturer, has been interpolated by a polynomial and fitted to the pixel versus photon-energy points. We also verify that the dispersion characteristic of the Wollaston polarizer is similar for the two polarizations. The bandwidth-per-pixel is $\sim$0.15 nm in the UV region to $\sim$3 nm in the IR region, resulting in an effective resolution of the order $\sim$1 nm to $\sim$20 nm respectively, considering a spot size of 6-7 pixels on the detectors.

\begin{figure*}[htbp]
\centering
\fbox{\includegraphics[width=0.8\linewidth]{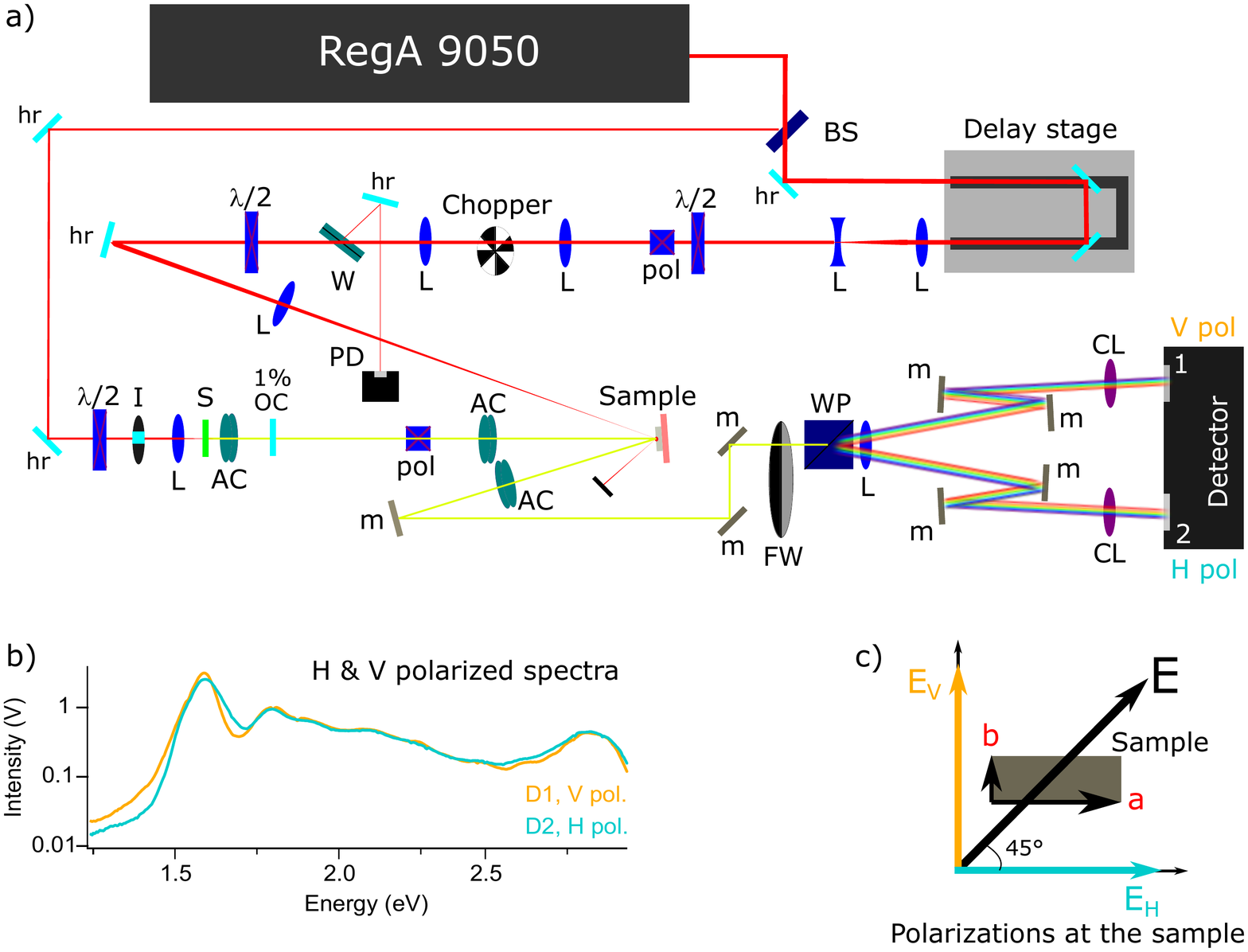}}
\caption{a) Sketch of the setup for polarization-resolved broadband time-resolved optical spectroscopy. The broadband supercontinuum probe beam is linearly polarized at 45 degrees with respect to the optical table. After interaction with the sample, the beam is split and dispersed on the detectors by a Wollaston polarizer. The labels of the optical elements have the following meaning: L=lens; AC=achromatic doublet; hr=800 nm dielectric mirror; m=Ag metallic mirror; I=iris; W=window; S=sapphire crystal; 1$\%$ OC=output coupler with T=1$\%$ at 800 nm; FW=filter wheel; WP=Wollaston polarizer; CL=cylindrical lens. b) H and V spectra as measured by the two detectors. c) Geometry of the experiment: the vertically-polarized component interacts with the b axis of the sample and is detected by detector '1'; the horizontally-polarized component interacts with the a axis of the sample and is detected by detector '2'. }
\label{fig1}
\end{figure*}

A home-made detector based on two Hamamatsu S10453 linear photodiodes array detectors, digitized with a 16 bit - 10 MHz digitizer, is used for simultaneous detection of the two beams. The detectors have 512 pixels, each 25 $\mu$m wide. The maximum line-rate achievable is $\sim$18 kHz. The integration time of the detectors is maximized to match this rate, hence every acquired spectrum is the average of 14 laser pulses. The same digitizer samples synchronously the pump beam status, modulated by a mechanical chopper at $\approx$4 kHz (the pump beam is focused on the chopper blade by a telescope). The setup can operate alternatively in reflection or in transmission. For the transmission geometry, the probe beam hits the sample at normal incidence. For the reflection geometry, slightly off-normal incidence is used (incidence angle $\sim$2.5 degrees) for collecting the reflected beam. This approach permits to use the normal-incidence expressions for the reflectivity or transmittivity when modelling the data in terms of a dielectric function. In addition, both probe polarization components can be assumed to be $s$-polarized. The pump beam hits the sample at an incidence angle of $\sim$15 degrees. Its polarization direction is set to 45 degrees with respect to the horizontal (and orthogonal to the probe polarization). In this way, the pump beam will not excite preferentially the sample along one specific crystalline direction. In the design of the setup, transmittive optical elements at normal incidence have been used in order to introduce negligible polarization anisotropies (see Fig. 1); the drawback is the introduction of a large chirp in the pulse because of the dispersion of its spectral components while passing through the optical elements. However, as demonstrated in \cite{Polli2010}, multichannel detection allows one to recover a high temporal resolution. The normalized reflectivity variation maps, $\Delta R/R(t,h\nu)$, are shown in this paper after a correction of the chirp.
We also developed a setup for ultra-broadband time-resolved optical spectroscopy, with which we acquire complementary measurements. Its distinctive feature is the ultra-broadband probing capability, covering a bandwidth in excess of 1000 nm, in the range 470-1550 nm (corresponding to 0.8-2.7 eV). This result is achieved thanks to the use of InGaAs detectors (Hamamatsu G11608-256) with special doping, that extends their sensitivity toward the visible spectral range. In order to achieve a sizeable dispersion in the near-IR range, we make use of an equilateral SF11 prism, that renders this scheme best suited to operate in $p$ polarization.

\section{Transient Reflectivity of M\lowercase{O}T\lowercase{E}$_2$ Polytypes}
We perform time-resolved reflectivity measurements on the molybdenum ditelluride (MoTe$_2$) system. It can be synthesized under the form of different polymorphs, differing for both the electronic and crystal structure (and their type of symmetry). In particular, the possible polytypes are three: the 2H form (an indirect-gap semiconductor with hexagonal lattice), the 1T' form and the Td form. The last two structures share a semimetallic electronic structure, but differ in the crystal structure, which is monoclinic for the 1T' form and orthorhombic for the Td form \cite{Dawson1987,Naylor2016}. A phase transition among these last two structures is obtained by varying the temperature $T$ across $T_c\sim$240 K, from the 1T' high-$T$ phase to the Td low-$T$ phase \cite{Crepaldi2017}. MoTe$_2$ gained recently large attention for a number of important properties, including a Type-II Weyl-semimetal character (in the Td non-centrosymmetric phase) \cite{Deng2016,Jiang2017,Soluyanov2015}, an extremely large magnetoresistance \cite{Chen2016b,Yang2017,Lee2018}, the possibility to easily obtain monolayers, displaying modified electronic properties with respect to their bulk counterpart (the 2H phase becomes a direct-gap semiconductor) \cite{Jariwala2014,Keum2015}, and finally, for the possibility of inducing phase transitions among different polytypes \cite{Zhang2019,Wang2019,Huang2016,Hwang2017}, including topological phase transitions. We perform measurements at room-temperature ($T$=300 K), hence we probe the 2H and 1T' phases. Both samples (provided by HQ Graphene) have been aligned with the a (x) axis along the horizontal direction.

\begin{figure*}[htbp]
\centering
\fbox{\includegraphics[width=0.8\linewidth]{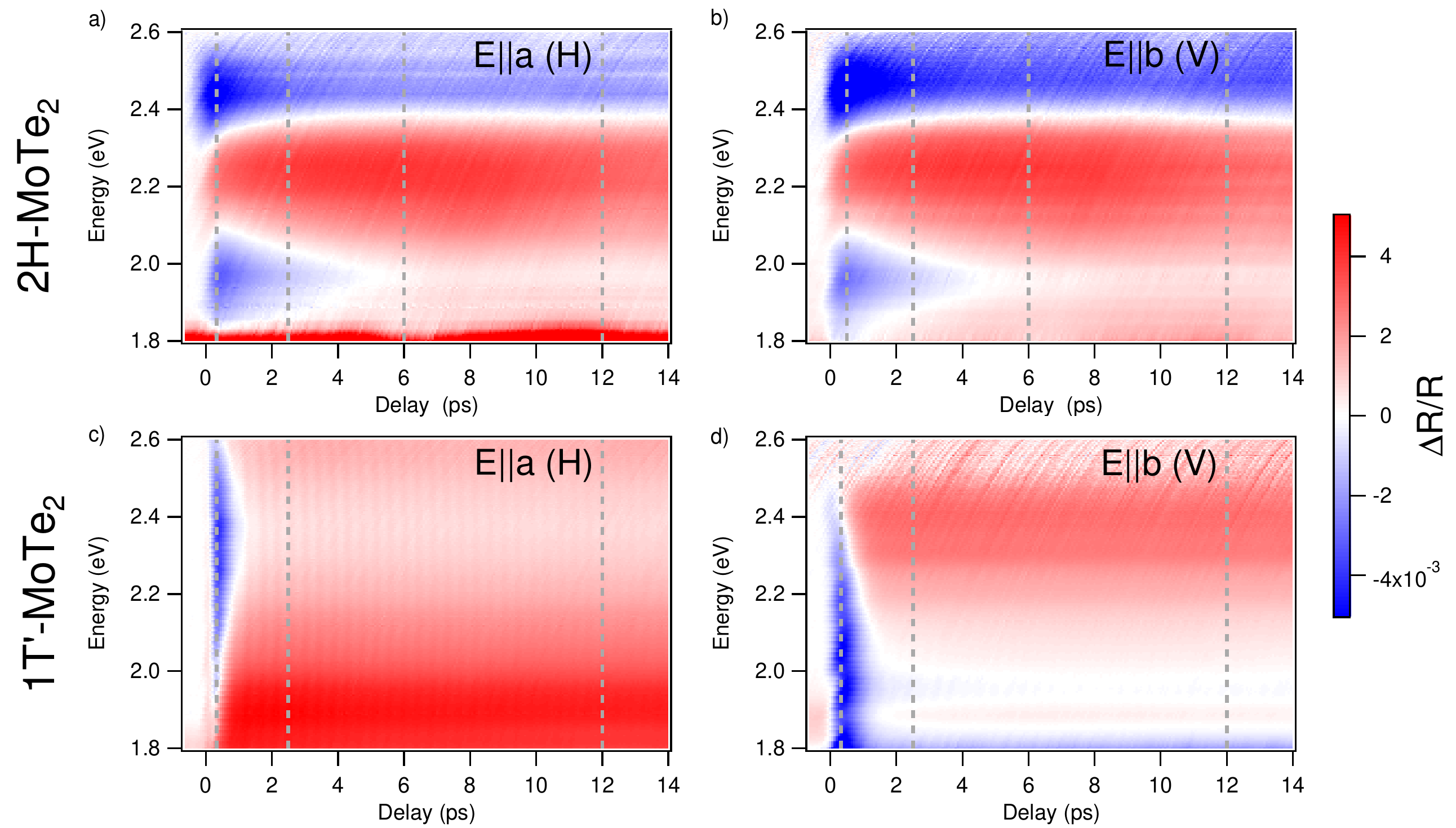}}
\caption{Maps of the time-and-spectrally resolved normalized reflectivity variations, $\Delta R/R(t,h\nu)$, for the polymorphic compounds 2H-MoTe$_2$ and 1T'-MoTe$_2$. The $\Delta R/R$ signal is displayed as a false-color map; the colorscale is indicated on the right. The probed crystallographic axis is indicated on top of each map. The hexagonal structure of the 2H polytype leads to isotropic non-equilibrium optical properties; the monoclinic structure of the 1T' polytype leads instead to markedly anisotropic non-equilibrium optical properties.}
\label{fig2}
\end{figure*}

Fig. 2 shows the $\Delta R/R(t,h\nu)$ maps for the 2H compound (panels a,b) and 1T' compound (panels c,d). They are displayed as a function of the pump-probe delay $t$ and probe photon energy h$\nu$, for both horizontal polarization H (probing the x-axis optical properties) and vertical polarization V (probing the y-axis optical properties). The pump fluence used for all the experiments is set to 400$\pm$50 $\mu$J/cm$^2$. The two compounds show markedly different time-resolved optical properties. More interestingly, a striking difference in their symmetry for H and V probing directions is revealed. The 2H polytype displays isotropic $\Delta R/R(t,h\nu)$ along the a and b axis, consistently with its uniaxial crystal structure. At variance, the 1T' polytype shows largely anisotropic signals, due to the biaxial crystalline structure. These results originate from the symmetry of the equilibrium optical properties, connected to the symmetry of each lattice system. In order to investigate in detail the time-resolved optical properties of the two compounds, we consider slices of the $\Delta R/R(t,h\nu)$ maps at fixed pump-probe delay and photon energy.

\begin{figure*}[htbp]
\centering
\fbox{\includegraphics[width=0.8\linewidth]{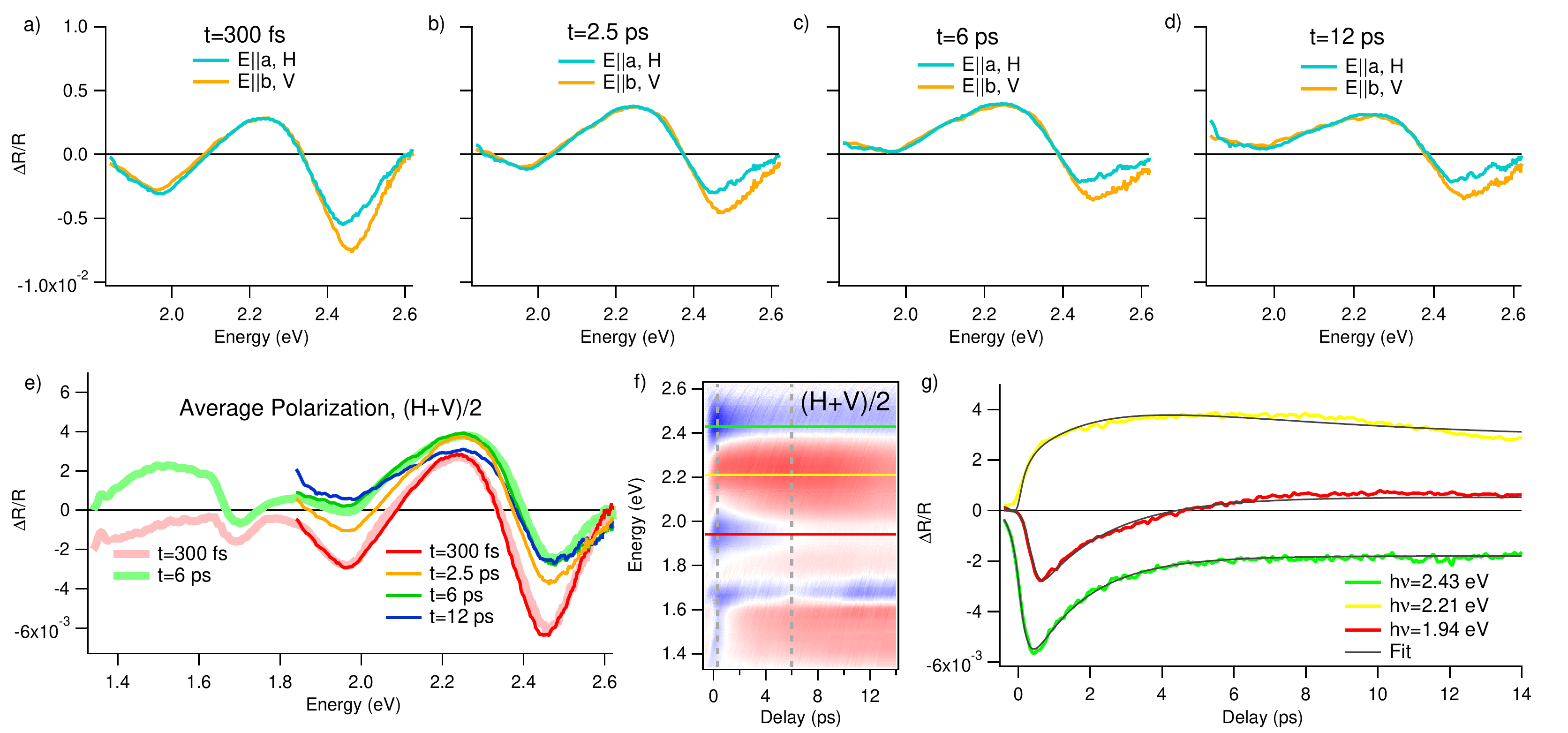}}
\caption{Spectral profiles for 2H-MoTe$_2$ at constant time-delays (indicated in Fig. 2a,b): a) $t$=300 fs; b) $t$=2.5 ps; c) $t$=6 ps; d) $t$=12 ps), for the a and b axis. At all delays, the time-resolved optical properties are isotropic. e) The average of H and V components in the range 1.8-2.6 eV is compared to a measurement on a larger (1.35-2.6 eV) spectral range (thick curves), obtained by placing a polarizer on the probe beam after interaction with the sample. The corresponding measurement is shown in f). Panel g) shows the relaxation dynamics as recorded at $h\nu$=2.43 eV, $h\nu$=2.21 eV, $h\nu$=1.94 eV. Black lines are the fit to the data, obtained as described in the main text.}
\label{fig3}
\end{figure*}

Fig. 3 shows the results recorded on 2H-MoTe$_2$. $\Delta R/R(t,h\nu)$ slices at fixed pump-probe delay ($t$=300 fs, 2.5 ps, 6 ps, 12 ps) are compared for H and V probe directions. They are reported in panels a)-d) of Fig. 3, respectively, to show to what extent our setup is capable of providing accurate measurements. In the case of 2H polytype, we indeed expect an isotropic behavior. By comparing the spectra for H and V polarization, it emerges that the setup provides accurate results in a wide spectral range, ranging from $\sim$1.8 eV to $\sim$2.4 eV. In the UV side we notice a small discrepancy that could be due to a non-perfect balancement of the H and V spectra. In the near-IR side instead, the results are affected by the scattering of the pump pulse at 800 nm (1.55 eV) on the detectors. This effect is attenuated by placing a polarizer on the reflected probe beam, for pump and probe beams are cross-polarized. This expedient allows us to obtain clean $\Delta R/R$ maps down to $\approx$1.35 eV, as shown in Fig. 3f. However, because of the polarizer, the possibility to measure anisotropies is lost, since each detector naturally samples an average signal equal to ($\Delta R/R_{H}$+$\Delta R/R_{V}$)/2$\equiv$(H+V)/2. We show in Fig. 3e the evolution of the $\Delta R/R(h\nu)$ signal for the selected delays. Curves have been obtained by averaging the H, V channels. The time-evolution shows a picosecond-timescale decay, with $\Delta R/R$ assuming more positive values on the full spectral range. We superimpose in Fig. 3e the data extracted from the dataset reported in Fig. 3f, covering a wider photon energy range (1.35-2.6 eV). The results are in very good agreement. Hence, at the expense of polarization sensibility, an extended spectral range is accessible. Finally, Fig. 3g shows the temporal dynamics extracted at three photon energies, $h\nu$=2.43 eV, $h\nu$=2.21 eV, $h\nu$=1.94 eV. A picoseconds-timescale relaxation dynamics is detected. Solid lines are the fit to the data obtained with a double-exponential decay convoluted with a Gaussian function having FWHM (Full-Width-at-Half-Maximum) of 100 fs, accounting for the experimental time resolution (pump-probe cross-correlation). At $h\nu$=2.43 eV and $h\nu$=1.94 eV, the dynamics can be captured by a single exponential component, with negative sign and relaxation timescale of 1.75$\pm$0.15 ps. Two exponential components are required instead to fit the dynamics at $h\nu$=2.21 eV. In addition to the negative component already discussed (with unvaried timescale), a positive component with exponential time-constant of 6.0$\pm$0.2 ps is required to properly fit the data. The two exponential components detected are likely associated to two different relaxations channels involving different pathways and optical transitions.

\begin{figure*}[htbp]
\centering
\fbox{\includegraphics[width=0.8\linewidth]{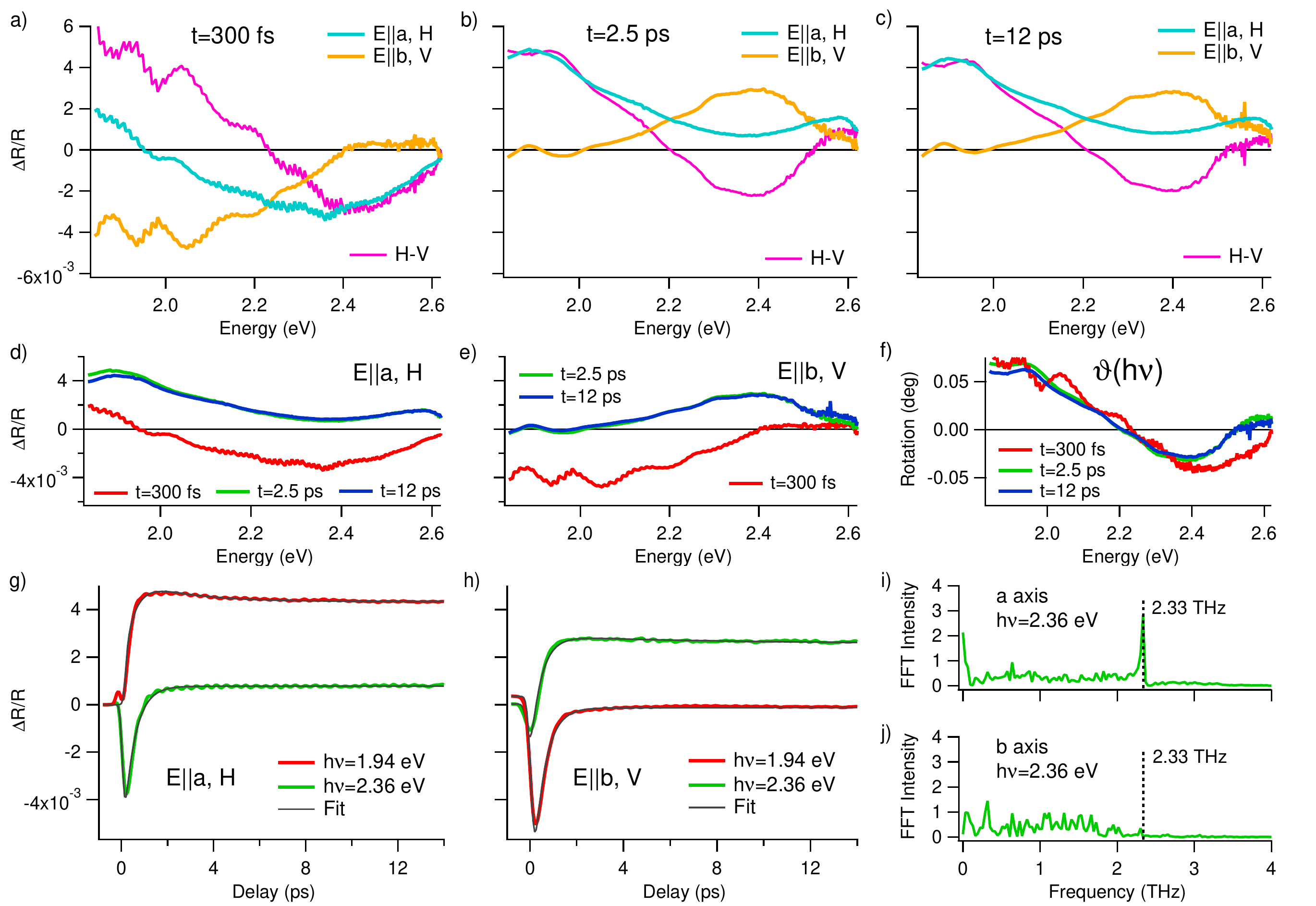}}
\caption{Spectral profiles for 1T'-MoTe$_2$ at constant time-delays (indicated in Fig. 2c,d): a) $t$=300 fs; b) $t$=2.5 ps; c) $t$=12 ps) for the a and b axis. The time-resolved optical properties here are markedly anisotropic, for all time-delays. In each panel we also show the difference among H and V signals. Panels d), e) show the time-evolution of the H and V spectra respectively. After a fast relaxation, the signal reaches a plateau lasting for several picoseconds. f) The anisotropy of the $\Delta R/R(h\nu)$ signal along the H and V axes induces an effective rotation of the polarization. We display the rotation as a function of the photon energy (see main text) at $t$=300 fs, $t$=2.5 ps and $t$=12 ps. g), h) The dynamics at $h\nu$=2.36 eV and $h\nu$=1.94 eV are shown, for the a and b axes respectively. Coherent phonon features are revealed predominantly in the a-axis curves. The black lines are the fit to the data, for the incoherent part of the signal. i), j) The Fast-Fourier-Transforms of the residuals at $h\nu$=2.36 eV show a clear coherent phonon feature at 2.33 THz, ascribed to the ${}^{1}A_{g}$ phonon mode (77 cm$^{-1}$). A large asymmetry of the coherent phonon effect along the a and b axes is revealed.}
\label{fig4}
\end{figure*}

The non-equilibrium optical properties recorded on the 1T' polytype present markedly different features, as reported in Fig. 4. Panels a), b), c) show slices at fixed pump-probe delay ($t$=300 fs, 2.5 ps, 12 ps) for both H and V probe directions, respectively. For all the time delays, a marked asymmetry is evident, on the whole photon energy range. In each panel we also plot the difference among the H and V components, to emphasize the anisotropy. Panels d) and e) show the evolution of the $\Delta R/R(h\nu)$ signal for H and V polarizations respectively, at the selected delays. In the case of 1T'-MoTe$_2$ compound, the spectra show a fast relaxation toward a plateau lasting for several tens of picoseconds. For both polarizations, the $\Delta R/R(h\nu)$ signal assumes more positive values at long pump-probe delays. Fig. 4f shows the rotation of the reflected probe linear polarization with respect to the incident direction, expressed in degrees. The expression we derived, valid for a balanced incident probe beam (that is, oriented at 45 degrees), is: $\theta=(1/4)(H-V)=(1/4)(\Delta R/R_{H}-\Delta R/R_{V})$, which holds for small rotation angles. The rotation arises from the fact that a different magnitude of the $\Delta R/R(h\nu)$ signal along two orthogonal directions, due to the anisotropic optical properties of the material, leads to an effective rotation of the direction of the electric field of the reflected beam, with respect to the incident one. In the case of 1T'-MoTe$_2$, this rotation has a strong photon energy dependence, with a maximum value of the order 0.05 degrees at $h\nu\sim$1.9 eV, and is almost constant in the range 1.8-2.4 eV at all delays. Fig. 4g and 4h show the relaxation dynamics as extracted at $h\nu$=2.36 eV and $h\nu$=1.94 eV, for H and V polarization respectively. The dynamics have been fitted by two exponential decays: a fast, negative component with time-constant $\tau_1$=460$\pm$20 fs, and a positive component with a larger timescale ($\tau_2\sim$2-8 ps) that brings the $\Delta R/R(t)$ signal to a plateau which lasts for times well in excess of the investigated delay range. The results of the fit are displayed as black solid lines. This dynamics is consistent with the metallic nature of the 1T'-MoTe$_2$ compound. The fast dynamics (described by $\tau_1$) can be associated to electron-phonon scattering processes; the slower timescale (described by $\tau_2$) is originated instead by the superposition of spectroscopic features of $\Delta R/R(t,h\nu)$ having opposite sign and different magnitude, giving rise to a fictitious dynamics. Finally, the long-lasting plateau is likely due to the heating of the lattice. Superimposed to the incoherent dynamics of 1T'-MoTe$_2$, we reveal periodic oscillations, indicating the excitation of coherent phonons. We analyze in detail the features at $h\nu$=2.36 eV; however, similar results are obtained at $h\nu$=1.94 eV. The Fast-Fourier-Transform (FFT) of the residuals (obtained by subtracting the result of the fittings discussed above to the time-traces) are reported in Fig. 4i), 4j) for H and V polarization respectively. A sharp feature at 2.33 THz (corresponding to 77.7 cm$^{-1}$) is revealed, in agreement with previous Raman studies \cite{Zhang2017,Ma2016,Oliver2017,Song2017,Chen2016a} reporting on a Raman-active ${}^{1}A_{g}$ phonon mode at 77 cm$^{-1}$. In our data, this phonon feature is much more pronounced in H polarization (that is, along the a axis) than in the orthogonal direction. This holds for the whole spectral range explored, hence, we can conclude that the ${}^{1}A_{g}$ phonon mode at $\sim$77 cm$^{-1}$ mainly affects the optical properties along the a-axis \cite{Ma2016}.

\section{Discussion}
 Measurements collected on the 1T'-MoTe$_2$ polytype prove the importance to measure the $\Delta R/R(t,h\nu)$ signal along both principal dielectric axis in the case of bi-axial crystals, representing a large number of the possible crystal structures. In our experiment, the time and spectral dependence of the time-resolved optical properties provides information about the ground-state electronic structure of the two polytypes: the semiconducting 2H compound displays a relaxation dynamics of the order 2 ps; the semi-metallic 1T' compound displays a faster (<500 fs) relaxation dynamics, and a plateau indicating the heating of the lattice system. In order to investigate more deeply the origin and the spectral dependence of the $\Delta R/R$ signal in 2H-MoTe$_2$, we make use of a setup for ultra-broadband time-resolved optical spectroscopy, covering the range 0.8-2.7 eV.
 
\begin{figure}[htbp]
\centering
\fbox{\includegraphics[width=0.7\linewidth]{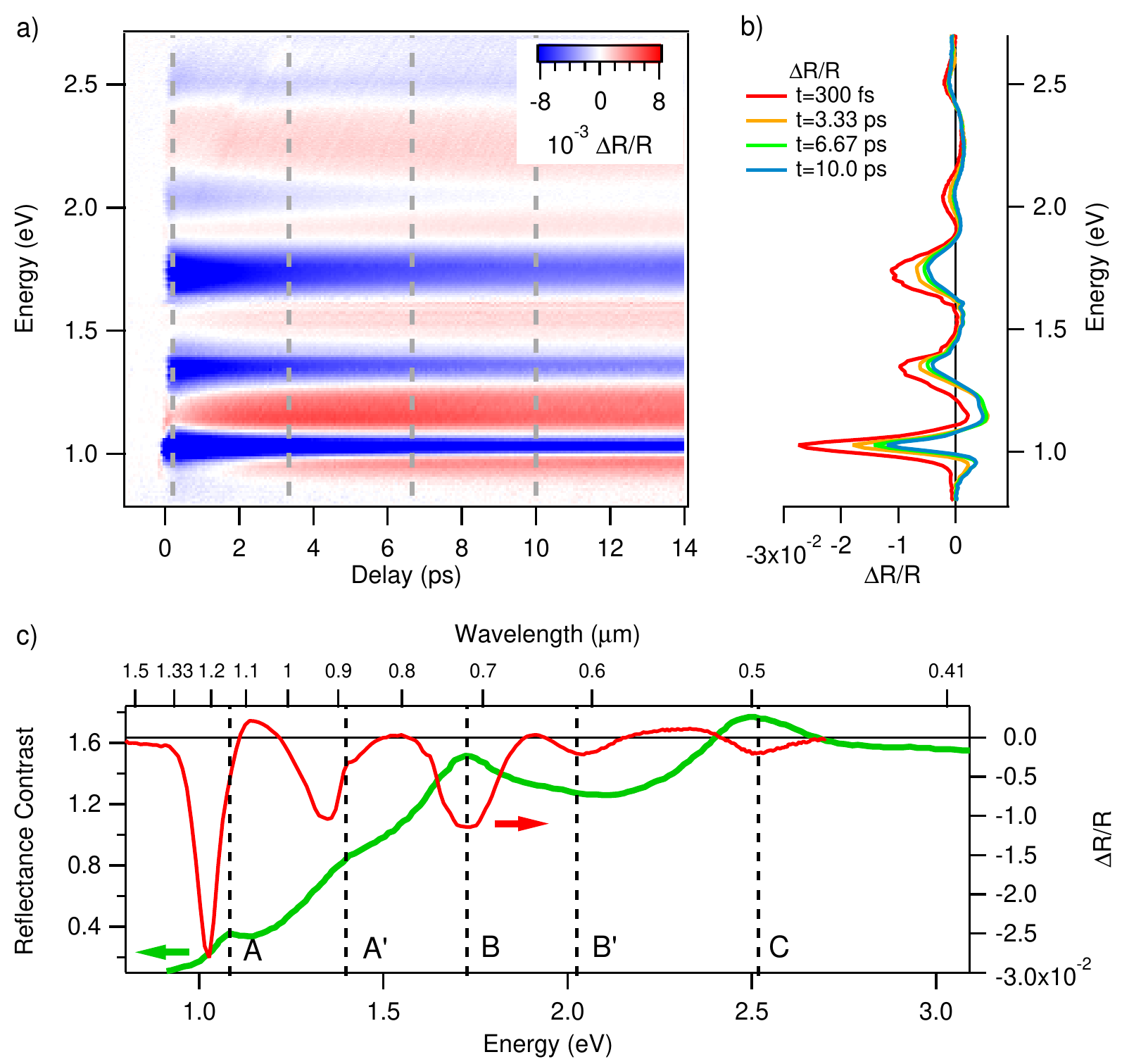}}
\caption{a) Time-resolved reflectivity measurements on 2H-MoTe$_2$, as recorded by an ultra-broadband setup extending the sensitivity to the near-infrared range. The large and sharp modulations are ascribed to the formation of excitons. b) The $\Delta R/R(h\nu)$ signal as collected at pump-probe delays $t$=300 fs, $t$=3.33 ps, $t$=6.67 ps, $t$=10.0 ps, marked in panel a), is shown. c) The signal at $t$=300 fs (red curve) is compared to the results of reflectance contrast measurements (green curve) taken from \cite{Ruppert2014}, showing excellent agreement and demonstrating the possibility to measure simultaneously five exciton features with an out-of-equilibrium experiment.}
\label{fig5}
\end{figure}
 
 Fig. 5 shows the results obtained with this apparatus. The probe polarization is oriented along the x axis, and the pump fluence is set to 400$\pm$50 $\mu$J/cm$^2$. Fig. 5a displays the $\Delta R/R(t,h\nu)$ map, revealing a marked modulation consisting of a sequence of very sharp spectral features. Cuts at constant pump-probe delays ($t$=300 fs, 3.33 ps, 6.67 ps, 10.0 ps) are reported in Fig. 5b. The main spectral features are in agreement with those reported in Fig. 3; on a few-picoseconds timescale, $\Delta R/R(h\nu)$ assumes more positive values and gets smaller in amplitude. The signal becomes more pronounced in the infrared spectral range, with a marked peak at $\sim$1 eV (1200 nm). Fig. 5c shows a comparison of the spectrum collected at $t$=300 fs with the result of reflectance contrast measurements reported in \cite{Ruppert2014}. This comparison shows that the sharp spectral features can be unambiguously associated to transitions of excitonic nature (termed A, A', B, B', C), associated to optical transitions between the top of the valence band and the bottom of the conduction band at the K point of the Brillouin zone \cite{Ramasubramaniam2012}. A very good agreement between the exciton energies for the three-layer compound studied in \cite{Ruppert2014} and the locations of the minima of the $\Delta R/R(h\nu)$ signal we measured on a bulk sample is obtained. The results are also in excellent agreement with the calculations from first principles reported in \cite{Ramasubramaniam2012}. Remarkably, the ultra-broadband probe spectral range of the setup allows us to disclose in a single non-equilibrium experiment all the five excitonic transitions of 2H-MoTe$_2$. Similar features have been revealed in a time-resolved absorption experiment on single-layer MoS$_2$ \cite{Pogna2016}. The transients recorded across excitonic transitions of single-layer MoS$_2$ share a close spectral dependence of those we record on the bulk 2H-MoTe$_2$ sample. The intense and negative signal was associated to a reduced absorption at the excitonic resonance, while the weaker positive signal at slightly lower photon energy to a red-shifted photoinduced absorption. This sequence repeats at all excitonic transitions, and was associated to a photoinduced transient bandgap renormalization caused by the presence of photoexcited carriers \cite{Pogna2016}, that affects simultaneously all of the five excitonic peaks. With the time resolution of the present experiments, we cannot reveal any difference in the exciton build-up time.
 
\section{Conclusions}
In conclusion, we demonstrated that the possibility to perform simultaneously TR-OS measurements along two principal dielectric axes of biaxial crystalline compounds can provide important information about the symmetry of their crystal structure. We carried out systematic time-resolved reflectivity experiments on two polytypes of MoTe$_2$, the semiconducting 2H phase and the semimetallic 1T' phase. In the case of the 1T' structure, marked anisotropies in the time-resolved optical properties are revealed, as well as in the coherent phonon features. These results can advance the comprehension of the out-of-equilibrium electronic and lattice dynamics of anisotropic materials, providing more constraints to data modelling and more solid bases to the interpretation. The possibility to obtain a photon-energy-dependent rotation of the polarization direction of the reflected beam could be used to engineer devices capable to rotate the polarization of a linearly polarized beam on ultrafast timescales. In the case of the 2H semiconducting structure, an ultra-broadband probe with >1000 nm bandwidth allowed us to evidence spectral features connected to five excitons, whose energies are in agreement with the result of independent measurements.
In perspective, our setup can find important applications in the study of a wide class of biaxial crystals or for systems where the anisotropy is emerging as a consequence of purely electronic effects. One important example is constituted by the electronic nematic phase \cite{Luo2017}, in which a compound enters when cooled down below an activation temperature. This effect is usually associated to a differentiation of the occupation of otherwise degenerate electronic states. Another application is the study of polarization rotations arising from the valley dynamics in single-layer transition-metal dichalcogenides. Our approach can find application also for the study and control of photoinduced phase transitions among different crystal structures. Recent examples are the ultrafast topological phase transition between Td and the 1T' structures of MoTe$_2$ \cite{Zhang2019} and WTe$_2$ \cite{Edbert2019}, or a reversible phase transition between the 2H and 1T' structures of monolayer MoTe$_2$, which has been demonstrated by means of ionic liquid gating \cite{Wang2019}. Finally, our development can find applications for the measurement of spectrally-resolved polarization rotations, originated by both magnetic and non-magnetic compounds.

\section{Disclosures}
The authors declare no conflicts of interest.

\bibliography{Main}

\begin{thebibliography}{10}
\newcommand{\enquote}[1]{``#1''}

\bibitem{Giannetti_Review}
C.~Giannetti, M.~Capone, D.~Fausti, M.~Fabrizio, F.~Parmigiani, and
  D.~Mihailovic, \enquote{Ultrafast optical spectroscopy of strongly correlated
  materials and high-temperature superconductors: a non-equilibrium approach,}
  {\protect\JournalTitle{Advances in Physics}} \textbf{65}, 58--238 (2016).

\bibitem{Liu2008}
Y.~H. Liu, Y.~Toda, K.~Shimatake, N.~Momono, M.~Oda, and M.~Ido,
  \enquote{Direct observation of the coexistence of the pseudogap and
  superconducting quasiparticles in {B}i$_2${S}r$_2${C}a{C}u$_2${O}$_{8+y}$ by
  time-resolved optical spectroscopy,} {\protect\JournalTitle{Phys. Rev.
  Lett.}} \textbf{101}, 137003 (2008).

\bibitem{Tomeljak2009}
A.~Tomeljak, H.~Schäfer, D.~Städter, M.~Beyer, K.~Biljakovic, and J.~Demsar,
  \enquote{Dynamics of photoinduced charge-density-wave to metal phase
  transition in {K}$_{0.3}${M}o{O}$_3$,} {\protect\JournalTitle{Phys. Rev.
  Lett.}} \textbf{102}, 066404 (2009).

\bibitem{Polli2007}
D.~Polli, L.~Lüer, and G.~Cerullo, \enquote{High-time-resolution pump-probe
  system with broadband detection for the study of time-domain vibrational
  dynamics,} {\protect\JournalTitle{Review of Scientific Instruments}}
  \textbf{78}, 103108 (2007).

\bibitem{Leonard2007}
J.~Léonard, N.~Lecong, J.-P. Likforman, O.~Crégut, S.~Haacke, P.~Viale,
  P.~Leproux, and V.~Couderc, \enquote{Broadband ultrafast spectroscopy using a
  photonic crystal fiber: application to the photophysics of malachite green,}
  {\protect\JournalTitle{Optics Express}} \textbf{15}, 16124--16129 (2007).

\bibitem{Cilento2010}
F.~Cilento, C.~Giannetti, G.~Ferrini, S.~D. Conte, T.~Sala, G.~Coslovich,
  M.~Rini, A.~Cavalleri, and F.~Parmigiani, \enquote{Ultrafast
  insulator-to-metal phase transition as a switch to measure the spectrogram of
  a supercontinuum light pulse,} {\protect\JournalTitle{Applied Physics
  Letters}} \textbf{96}, 021102 (2010).

\bibitem{Chergui2012a}
G.~Auböck, C.~Consani, R.~Monni, A.~Cannizzo, F.~van Mourik, and M.~Chergui,
  \enquote{Femtosecond pump/supercontinuum-probe setup with 20 k{H}z repetition
  rate,} {\protect\JournalTitle{Review of Scientific Instruments}} \textbf{83},
  093105 (2012).

\bibitem{Megerle2009}
U.~Megerle, I.~Pugliesi, C.~Schriever, C.~F. Sailer, and E.~Riedle,
  \enquote{Sub-50 fs broadband absorption spectroscopy with tunable excitation:
  putting the analysis of ultrafast molecular dynamics on solid ground,}
  {\protect\JournalTitle{Appl. Phys. B}} \textbf{96}, 215 (2009).

\bibitem{Giannetti2011}
C.~Giannetti, F.~Cilento, S.~D. Conte, G.~Coslovich, G.~Ferrini, H.~Molegraaf,
  M.~Raichle, R.~Liang, H.~Eisaki, M.~Greven, A.~Damascelli, D.~van~der Marel,
  and F.~Parmigiani, \enquote{Revealing the high-energy electronic excitations
  underlying the onset of high-temperature superconductivity in cuprates,}
  {\protect\JournalTitle{Nature Communications}} \textbf{2}, 353 (2011).

\bibitem{Preda2016}
F.~Preda, V.~Kumar, F.~Crisafi, D.~G. F.~D. Valle, G.~Cerullo, and D.~Polli,
  \enquote{Broadband pump-probe spectroscopy at 20-{MH}z modulation frequency,}
  {\protect\JournalTitle{Optics Letters}} \textbf{41}, 2970 (2016).

\bibitem{Chergui2012b}
G.~Auböck, C.~Consani, F.~van Mourik, and M.~Chergui, \enquote{Ultrabroadband
  femtosecond two-dimensional ultraviolet transient absorption,}
  {\protect\JournalTitle{Optics Letters}} \textbf{37}, 2337 (2012).

\bibitem{Hiramatsu2015}
K.~Hiramatsu and T.~Nagata, \enquote{Communication: Broadband and
  ultrasensitive femtosecond time-resolved circular dichroism spectroscopy,}
  {\protect\JournalTitle{J. Chem. Phys}} \textbf{143}, 121102 (2015).

\bibitem{Wilson1969}
J.~A. Wilson and A.~D. Yoffe, \enquote{The transition metal dichalcogenides
  discussion and interpretation of the observed optical, electrical and
  structural properties,} {\protect\JournalTitle{Advances in Physics}}
  \textbf{18}, 193--335 (1969).

\bibitem{BOYD}
R.~W. Boyd, \enquote{Nonlinear optics (third edition),} in \emph{Nonlinear
  Optics (Third Edition),}  (Elsevier Academic, 2007).

\bibitem{Bradler2009}
M.~Bradler, P.~Baum, and E.~Riedle, \enquote{Femtosecond continuum generation
  in bulk laser host materials with sub-$\mu$j pump pulses,}
  {\protect\JournalTitle{Appl. Phys. B}} \textbf{97}, 561–574 (2009).

\bibitem{Lu2014}
C.-H. Lu, Y.-J. Tsou, H.-Y. Chen, B.-H. Chen, Y.-C. Cheng, S.-D. Yang, M.-C.
  Chen, C.-C. Hsu, and A.~H. Kung, \enquote{Generation of intense
  supercontinuum in condensed media,} {\protect\JournalTitle{Optica}}
  \textbf{1}, 400--406 (2014).

\bibitem{Dubietis2017}
A.~Dubietis, G.~Tamosauskas, R.~Suminas, V.~Jukna, and A.~Couairon,
  \enquote{Ultrafast supercontinuum generation in bulk condensed media,}
  {\protect\JournalTitle{Lituanian Journal of Physics}} \textbf{57}, 113--157
  (2017).

\bibitem{Polli2010}
D.~Polli, D.~Brida, S.~Mukamel, G.~Lanzani, and G.~Cerullo, \enquote{Effective
  temporal resolution in pump-probe spectroscopy with strongly chirped pulses,}
  {\protect\JournalTitle{Phys. Rev. A}} \textbf{82}, 053809 (2010).

\bibitem{Dawson1987}
W.~G. Dawson and D.~W. Bullett, \enquote{Electronic structure and
  crystallography of {M}o{T}e$_2$ and {W}{T}e$_2$,}
  {\protect\JournalTitle{Journal of Physics C: Solid State Physics}}
  \textbf{20}, 6159 (1987).

\bibitem{Naylor2016}
C.~H. Naylor, W.~M. Parkin, J.~Ping, Z.~Gao, Y.~R. Zhou, Y.~Kim, F.~Streller,
  R.~W. Carpick, A.~M. Rappe, M.~Drndić, J.~M. Kikkawa, and A.~T.~C. Johnson,
  \enquote{Monolayer single-crystal 1{T}'-{M}o{T}e$_2$ grown by chemical vapor
  deposition exhibits weak antilocalization effect,}
  {\protect\JournalTitle{Nano Lett.}} \textbf{16}, 4297--4304 (2016).

\bibitem{Crepaldi2017}
A.~Crepaldi, G.~Autes, A.~Sterzi, G.~Manzoni, M.~Zacchigna, F.~Cilento,
  I.~Vobornik, J.~Fujii, P.~Bugnon, A.~Magrez, H.~Berger, F.~Parmigiani, O.~V.
  Yazyev, and M.~Grioni, \enquote{Persistence of a surface state arc in the
  topologically trivial phase of {M}o{T}e$_2$,} {\protect\JournalTitle{Physical
  Review B}} \textbf{95}, 041408(R) (2017).

\bibitem{Deng2016}
K.~Deng, G.~Wan, P.~Deng, K.~Zhang, S.~Ding, E.~Wang, M.~Yan, H.~Huang,
  H.~Zhang, Z.~Xu, J.~Denlinger, A.~Fedorov, H.~Yang, W.~Duan, H.~Yao, Y.~Wu,
  S.~Fan, H.~Zhang, X.~Chen, and S.~Zhou, \enquote{Experimental observation of
  topological fermi arcs in type-{II} {W}eyl semimetal {M}o{T}e$_2$,}
  {\protect\JournalTitle{Nature Physics}} \textbf{12}, 1105–1110 (2016).

\bibitem{Jiang2017}
J.~Jiang, Z.~K. Liu, Y.~Sun, H.~F. Yang, C.~R. Rajamathi, Y.~P. Qi, L.~X. Yang,
  C.~Chen, H.~Peng, C.-C. Hwang, S.~Sun, S.-K. Mo, I.~Vobornik, J.~Fujii,
  S.~Parkin, C.~Felser, B.~Yan, and Y.~L. Chen, \enquote{Signature of type-{II}
  {W}eyl semimetal phase in {M}o{T}e$_2$,} {\protect\JournalTitle{Nature
  Communications}} \textbf{8} (2017).

\bibitem{Soluyanov2015}
A.~A. Soluyanov, D.~Gresch, Z.~Wang, Q.~Wu, M.~Troyer, X.~Dai, and B.~A.
  Bernevig, \enquote{Type-{II} {W}eyl semimetals,}
  {\protect\JournalTitle{Nature}} \textbf{527}, 495 (2015).

\bibitem{Chen2016b}
F.~C. Chen, H.~Y. Lv, X.~Luo, W.~J. Lu, Q.~L. Pei, G.~T. Lin, Y.~Y. Han, X.~B.
  Zhu, W.~H. Song, and Y.~P. Sun, \enquote{Extremely large magnetoresistance in
  the type-{II} {W}eyl semimetal {M}o{T}e$_2$,} {\protect\JournalTitle{Phys.
  Rev. B}} \textbf{94}, 235154 (2016).

\bibitem{Yang2017}
J.~Yang, J.~Colen, J.~Liu, M.~C. Nguyen, G.~wei Chern, and D.~Louca,
  \enquote{Elastic and electronic tuning of magnetoresistance in {M}o{T}e$_2$,}
  {\protect\JournalTitle{Science Advances}} \textbf{3} (2017).

\bibitem{Lee2018}
S.~Lee, J.~Jang, S.-I. Kim1, S.-G. Jung, J.~Kim, S.~Cho, S.~W. Kim, J.~Y. Rhee,
  K.-S. Park, and T.~Park, \enquote{Origin of extremely large magnetoresistance
  in the candidate type-{II} {W}eyl semimetal {M}o{T}e$_{2-x}$,}
  {\protect\JournalTitle{Scientific Reports}} \textbf{8} (2018).

\bibitem{Jariwala2014}
D.~Jariwala, V.~K. Sangwan, L.~J. Lauhon, T.~J. Marks, and M.~C. Hersam,
  \enquote{Emerging device applications for semiconducting two-dimensional
  transition metal dichalcogenides,} {\protect\JournalTitle{ACS Nano}}
  \textbf{8}, 1102--1120 (2014).

\bibitem{Keum2015}
D.~H. Keum, S.~Cho, J.~H. Kim, D.-H. Choe, H.-J. Sung, M.~Kan, H.~Kang, J.-Y.
  Hwang, S.~W. Kim, H.~Yang, K.~J. Chang, and Y.~H. Lee, \enquote{Bandgap
  opening in few-layered monoclinic {M}o{T}e$_2$,}
  {\protect\JournalTitle{Nature Physics}} \textbf{11}, 482–486 (2015).

\bibitem{Zhang2019}
M.~Y. Zhang, Z.~X. Wang, Y.~N. Li, L.~Y. Shi, D.~Wu, T.~Lin, S.~J. Zhang, Y.~Q.
  Liu, Q.~M. Liu, J.~Wang, T.~Dong, and N.~L. Wang, \enquote{Light-induced
  subpicosecond lattice symmetry switch in {M}o{T}e$_2$,}
  {\protect\JournalTitle{Physical Review X}} \textbf{9}, 021036 (2019).

\bibitem{Wang2019}
Y.~Wang, J.~Xiao, H.~Zhu, Y.~Li, Y.~Alsaid, K.~Y. Fong, Y.~Zhou, S.~Wang,
  W.~Shi, Y.~Wang, A.~Zettl, E.~J. Reed, and X.~Zhang, \enquote{Structural
  phase transition in monolayer {M}o{T}e$_2$ driven by electrostatic doping,}
  {\protect\JournalTitle{Nature}} \textbf{550}, 487 (2017).

\bibitem{Huang2016}
H.~H. Huang, X.~Fan, D.~J. Singh, H.~Chen, Q.~Jianga, and W.~T. Zheng,
  \enquote{Controlling phase transition for single-layer {M}{T}e$_2$ ({M} =
  {M}o and {W}): modulation of the potential barrier under strain,}
  {\protect\JournalTitle{Phys. Chem. Chem. Phys.}} \textbf{18}, 4086 (2016).

\bibitem{Hwang2017}
J.~Hwang, C.~Zhang, and K.~Cho, \enquote{Structural and electronic phase
  transitions of {M}o{T}e$_2$ induced by {L}i ionic gating,}
  {\protect\JournalTitle{2D Mater.}} \textbf{4}, 045012 (2017).

\bibitem{Zhang2017}
K.~Zhang, C.~Bao, Q.~Gu, X.~Ren, H.~Zhang, K.~Deng, Y.~Wu, Y.~Li, J.~Feng, and
  S.~Zhou, \enquote{Raman signatures of inversion symmetry breaking and
  structural phase transition in type-{II} {W}eyl semimetal {M}o{T}e$_2$,}
  {\protect\JournalTitle{Nature Communications}} \textbf{7} (2016).

\bibitem{Ma2016}
X.~Ma, P.~Guo, C.~Yi, Q.~Yu, A.~Zhang, J.~Ji, Y.~Tian, F.~Jin, Y.~Wang, K.~Liu,
  T.~Xia, Y.~Shi, , and Q.~Zhang, \enquote{Raman scattering in the
  transition-metal dichalcogenides of 1{T}'-{M}o{T}e$_2$, {T}d-{M}o{T}e$_2$,
  and {T}d-{W}{T}e$_2$,} {\protect\JournalTitle{Phys. Rev. B}} \textbf{94},
  214105 (2016).

\bibitem{Oliver2017}
S.~M. Oliver, R.~Beams, S.~Krylyuk, I.~Kalish, A.~K. Singh, A.~Bruma,
  F.~Tavazza, J.~Joshi, I.~R. Stone, S.~J. Stranick, A.~V. Davydov, and P.~M.
  Vora, \enquote{The structural phases and vibrational properties of
  {M}o$_{1-x}${W}$_x${T}e$_2$ alloys,} {\protect\JournalTitle{2D Mater.}}
  \textbf{4} (2017).

\bibitem{Song2017}
Q.~Song, H.~Wang, X.~Pan, X.~Xu, Y.~Wang, Y.~Li, F.~Song, X.~Wan, Y.~Ye, and
  L.~Dai, \enquote{Anomalous in-plane anisotropic raman response of monoclinic
  semimetal 1{T}'{M}o{T}e$_2$,} {\protect\JournalTitle{Scientific Reports}}
  \textbf{7} (2017).

\bibitem{Chen2016a}
S.-Y. Chen, T.~Goldstein, D.~Venkataraman, A.~Ramasubramaniam, and J.~Yan,
  \enquote{Activation of new raman modes by inversion symmetry breaking in type
  {II} {W}eyl semimetal candidate {T}'‑{M}o{T}e$_2$,}
  {\protect\JournalTitle{Nano Lett.}} \textbf{16}, 5852--5860 (2016).

\bibitem{Ruppert2014}
C.~Ruppert, O.~B. Aslan, and T.~F. Heinz, \enquote{Optical properties and band
  gap of single- and few-layer {M}o{T}e$_2$ crystals,}
  {\protect\JournalTitle{Nano Lett.}} \textbf{14}, 6231--6236 (2014).

\bibitem{Ramasubramaniam2012}
A.~Ramasubramaniam, \enquote{Large excitonic effects in monolayers of
  molybdenum and tungsten dichalcogenides,} {\protect\JournalTitle{Phys. Rev.
  B}} \textbf{86}, 115409 (2012).

\bibitem{Pogna2016}
E.~A.~A. Pogna, M.~Marsili, D.~D. Fazio, S.~D. Conte, C.~Manzoni, D.~Sangalli,
  D.~Yoon, A.~Lombardo, A.~C. Ferrari, A.~Marini, G.~Cerullo, and D.~Prezzi,
  \enquote{Photo-induced bandgap renormalization governs the ultrafast response
  of {S}ingle-{L}ayer {M}o{S}$_2$,} {\protect\JournalTitle{ACS Nano}}
  \textbf{10}, 1182 (2016).

\bibitem{Luo2017}
C.-W. Luo, P.~C. Cheng, S.-H. Wang, J.-C. Chiang, J.-Y. Lin, K.-H. Wu, J.-Y.
  Juang, D.~A. Chareev, O.~S. Volkova, and A.~N. Vasiliev, \enquote{Unveiling
  the hidden nematicity and spin subsystem in {F}e{S}e,}
  {\protect\JournalTitle{Quantum Materials}} \textbf{2}, 1 (2017).

\bibitem{Edbert2019}
E.~J. Sie, C.~M. Nyby, C.~D. Pemmaraju, S.~J. Park, X.~Shen, J.~Yang, M.~C.
  Hoffmann, B.~K. Ofori-Okai, R.~Li, A.~H. Reid, S.~Weathersby, E.~Mannebach,
  N.~Finney, D.~Rhodes, D.~Chenet, A.~Antony, L.~Balicas, J.~Hone, T.~P.
  Devereaux, T.~F. Heinz, X.~Wang, and A.~M. Lindenberg, \enquote{An ultrafast
  symmetry switch in a {W}eyl semimetal,} {\protect\JournalTitle{Nature}}
  \textbf{565}, 61 (2019).

\end{thebibliography}







\end{document}